\documentclass{emulateapj}

  \usepackage{amsmath}
  \usepackage[utf8]{inputenc}
  \usepackage{graphicx}
  \usepackage{natbib}
  \usepackage{units}
  \usepackage{subfigure}

\newcommand{\lra}[1]{{ \left( #1 \right) }}
\newcommand{\lrb}[1]{{ \left[ #1 \right] }}

\newcommand{\fil}[1]{\langle #1 \rangle}

\newcommand{\pd}[1]{\frac{\partial}{\partial #1}}
\newcommand{\ppd}[2]{\frac{\partial #2}{\partial #1}}
\newcommand{\pdd}[1]{\frac{\partial^2}{\partial #1^2}}
\newcommand{\fa}{\frac{1}{a}}
\renewcommand{\fh}{\frac{\dot{a}}{a}}

\newcommand{\alignold}{}
\let\alignold=\align
\let\endalignold=\endalign

\renewenvironment{align}{%
\footnotesize%
\abovedisplayskip=8pt%
\belowdisplayskip=8pt%
\alignold}
{\endalignold}

\shorttitle{Adaptively refined large eddy simulations of clusters}
\shortauthors{Maier et al.}

\begin{document}

\title{Adaptively refined large eddy simulations of a galaxy cluster: turbulence modeling and 
the physics of the intra-cluster medium}


\author{A.\ Maier\altaffilmark{1}, L.\ Iapichino\altaffilmark{2}, W.\ Schmidt\altaffilmark{1} 
and J.C.\ Niemeyer\altaffilmark{3}}

\altaffiltext{1}{Lehrstuhl f\"ur Astronomie, Universit\"at W\"urzburg, 
Am Hubland, D-97074 W\"urzburg, Germany}
\altaffiltext{2}{Zentrum f\"ur Astronomie der Universit\"at Heidelberg, 
Institut f\"ur Theoretische Astrophysik, Albert-Ueberle-Str 2, D-69120
Heidelberg, Germany}
\altaffiltext{3}{Institut f\"ur Astrophysik, Universit\"at
  G\"ottingen, Friedrich-Hund-Platz 1, D-37077 G\"ottingen, Germany}



\begin{abstract}
We present a numerical scheme for modelling unresolved turbulence in
cosmological adaptive mesh refinement codes. As a first
application, we study the evolution of turbulence in the intra-cluster medium
and in the core of a galaxy cluster. Simulations with and without
subgrid scale model are compared in detail. Since the flow in the ICM is subsonic,
the global turbulent energy
contribution at the unresolved length scales is smaller than $1 \%$ of
the internal energy. We find that the production of turbulence is
closely correlated with merger events occurring in the cluster
environment, and its dissipation locally affects the cluster energy
budget.  
Because of this additional source of dissipation, the core temperature
is larger and the density is smaller in the presence of subgrid scale turbulence
than in the standard adiabatic run, resulting in a higher entropy
core value.  
\end{abstract}


\section{Introduction}
\label{intro}
Simulations of cosmological structure formation often share two important attributes.
First, the ubiquitous presence of spatially localized features such as shocks,
clumps, or composition discontinuities that need to be numerically resolved or
at least adequately modeled; and second, moderate or large Reynolds numbers of the baryonic
component indicating that fully developed, i.e. space-filling turbulence is
responsible for the mixing and dissipation properties of the
gas. Despite great advances in computational fluid dynamics, an
accurate 
handling of both aspects has so far proven to be very difficult, because
dedicated numerical techniques seem to be mutually incompatible.

The most powerful technique for grid-based solvers to resolve localized
and anisotropic structures in a flow is adaptive mesh refinement (AMR) 
\citep{Berger1984,Berger1989}. This technique has proven to be very well
suited for several astrophysical problems \citep{n05}. However, in the
case of astrophysically relevant Reynolds numbers even with AMR we
cannot resolve all the relevant length scales down to the dissipative
one \citep{Schmidt2006}. Even if this condition may be achieved in
regions of maximum refinement, as is possibly the case in the core of
galaxy clusters (where the effective viscosity is still the subject of
debate), turbulence from coarser areas of the grid continuously flows
into these regions without being properly accounted for. 

In engineering applications as well as other fields of
computational fluid dynamics, subgrid scale (SGS) models have been
developed in order to mimic the influence of unresolved turbulence on
the resolved scales. This technique is often referred to as Large Eddy
Simulations (LES)
\citep{Lesieur1996}. In astrophysics, SGS models have already been
extensively used in simulations of
Type Ia supernova explosions \citep{nh95,rhn02,rh05,rhs07}.     
In this framework, \citet{Schmidt2006} presented a formulation of SGS models based on the
filtering approach of \citet{Germano1992}. Other applications of SGS
models in astrophysical problems have been proposed by
\citet{Pope2008} and, in an approach specially designed for
Rayleigh-Taylor-driven turbulence, by \citet{sb08}. 

In this paper, a numerical method that combines LES and AMR for the study of astrophysical turbulent 
flows will be presented. We will refer to this new tool as {\sc fearless} (Fluid mEchanics 
with Adaptively Refined Large Eddy SimulationS). With the combined use of grid refinement and SGS 
model, {\sc fearless} is very suitable for simulations of intermittent turbulent flows in clumped media. 

The formation and evolution of the cosmological large scale structure is a typical case of 
turbulence generation in a strongly clumped medium. 
The concordance model of cosmological structure formation
explains the formation of clusters through a hierarchical
sequence of mergers of lower-mass systems \citep[e.g.][]{Ostriker1993}. In particular, mergers of subhalos
play a fundamental role in 
determining the structure and dynamics of massive clusters of galaxies. Furthermore, it is known 
that major mergers induce temperature inhomogeneities and bulk motions
with velocities of the order of $\unit[1000]{km\ s^{-1}}$ in the intra-cluster medium (ICM)
\citep{Norman1999a}. This results in complex hydrodynamic flows where most of
the kinetic energy is quickly dissipated to heat by shocks, but some part may in principle also 
excite long-lasting turbulent gas motions. Besides merger processes, it is also known that galactic 
motions \citep{bd89,k07} and AGN outflows \citep{hby06,ss06a} can stir the ICM.

The problem of the turbulent state of the ICM is still
controversial, both from the theoretical point of view of constraining
the kinematic viscosity of the fluid \citep{rmf05,nm01,j08}, and from
the observational side, since a direct observation of turbulent
emission-line broadening is beyond the reach of current X-ray
observatories
\citep{Sunyaev2003,is03,bhr05,Dolag2005,rcs08}. Nonetheless, some
indirect ways of investigating turbulence in clusters gave encouraging
results (see \citealt{Iapichino2008} for an overview) and call for a
better theoretical understanding of the problem. A clear example for
the relevance of cluster turbulence for precision cosmology is
provided by recent results that demonstrate the sensitivity of
hydrostatic mass estimates on assumptions about the level of
turbulence \citep{lkn09}.

In numerical simulations of merging clusters \citep{Schindler1993, 
Roettiger1997,Ricker2001,fmw04,fsw04,Takizawa2005,ias08}, it has been shown that infalling
subclusters generate a laminar bulk flow but inject turbulent motions via
Kelvin-Helmholtz instabilities at the interfaces between the bulk flows and the
primary cluster gas. Such eddies redistribute the energy of the merger through
the cluster volume and decay into turbulent
velocity fields, eventually developing a turbulent cascade with a spectrum of
fluctuations expected to be close to a Kolmogorov spectrum
\citep{Dolag2005}. Numerical simulations focused on the role of
turbulence in astrophysical flows in general, and especially 
for clusters, have been restricted to measuring passively statistical quantities
like velocity dispersion from simulation data (e.g.~\citealt{Norman1999a,Dolag2005}). 
The active role of small scale velocity fluctuations on the large scale flow
has not been taken into account so far.

A previous attempt of modelling turbulence in hydrodynamical
simulation of cluster formation has been performed by
\citet{Iapichino2008}. In that work, the authors focused on better
definitions of the AMR criteria for refining the computational grid
where and when the flow in the ICM was turbulent
\citep{sfh08,ias08}. Though useful, this numerical strategy can follow
only a narrow range of large length scales along the turbulent
cascade, being the Kolmogorov length scale for turbulent dissipation
much lower than the spatial resolution. Besides this theoretical
shortcoming, also numerically it is questionable whether the mixing
forced at the mesh length scale correctly represents the physics of
turbulence \citep{mmb08}. 

These arguments motivate the application of {\sc fearless} to
cluster simulations as a more consistent approach. We show below that 
the additional degree of freedom given by the local turbulence
intensity on unresolved scales has a measurable impact on the
features of the ICM. In addition to the direct dynamical coupling to
the resolved fluid equations, the ability to separate unresolved
kinetic energy from thermal energy allows a more accurate computation
of the local temperature and entropy than without the subgrid scale
model. 

This work is structured as follows: in \S \ref{FGform}, the formalism
of the subgrid scale model and of {\sc fearless} is introduced. Some
numerical tests and consistency checks are presented in \S
\ref{tests}, and the setup of the galaxy cluster simulations is described
in \S \ref{details}. The results are presented in \S \ref{results} and
discussed in \S \ref{conclusions}, where our conclusions are drawn.

\section{Subgrid scale model and {\sc fearless}}\label{FGform}

\subsection{Germano decomposition}\label{decomp}

The dynamics of  a compressible, viscous, self-gravitating fluid 
with with density $\rho(r_i,t)$, momentum density $\rho v_i(r_i,t)$ and total
energy density $\rho e(r_i,t)$ at spatial position $(r_1,r_2,r_3)$ is given by the following set of equations:
\begin{align}
\pd{t}\rho + \pd{r_j}(v_j \rho) &= 0, \label{eq:mass}\\
\pd{t}(\rho v_i) + \pd{r_j}(v_j \rho v_i) &=
-\pd{r_i}p + \pd{r_j}\sigma'_{ij}+\rho g_i, \label{eq:mom} \\
\begin{split}
\pd{t}(\rho e) + \pd{r_j}(v_j \rho e) &=
-\pd{r_j}(v_j p) + \pd{r_j}(v_i \sigma'_{ij})+ v_i \rho g_i \,\,, \label{eq:etot}
\end{split}
\end{align}
where $p$ is the pressure, $g_i$ the gravitational acceleration and
$\sigma'_{ij}$ the viscous stress tensor. Note that the Einstein sum convention
applies to repeated indices. 

As shown by \citet{Schmidt2006}, these equations can be
decomposed into large-scale (resolved) and small-scale (unresolved)
parts using the filter formalism proposed by \citet{Germano1992}
in terms of density-weighted quantities\footnote{For a review, see \citet{RoepSchm09}.}.
By means of filtering, any field quantity $a$ can be split into a smoothed part
$\fil{a}$ and a fluctuating part $a'$, where $\fil{a}$
varies only at scales greater than the prescribed filter length.
We define density weighted filtered quantities according to \citet{Favre1969} by
\begin{align}
\fil{\rho a} = \fil{\rho} \hat{a} \Rightarrow \hat{a}=\frac{\fil{\rho
a}}{\fil{\rho}}\,\,.
\end{align}

Following \citet{Schmidt2006}, filtered equations for compressible fluid dynamics
can be derived:
\begin{align}
\pd{t}\fil{\rho} + \pd{r_j}\hat{v}_j\fil{\rho} =&\ 0, \label{eq:filmsum}\\
\scriptstyle
\begin{split}
\pd{t}\fil{\rho}\hat{v}_i + \pd{r_j}\hat{v}_j\fil{\rho}\hat{v}_i =&
-\pd{r_i}\fil{p}+\pd{r_j}\fil{\sigma'_{ij}}\\
&+\fil{\rho}\hat{g}_i-\pd{r_j}\hat{\tau}(v_i,v_j)\,\,,
\end{split}\label{eq:filmomsum}
\\
\begin{split}
\pd{t}\fil{\rho}e_{\mathrm{res}}+\pd{r_j}\hat{v}_j\fil{\rho}e_{\mathrm{res}}=&
-\pd{r_i}\hat{v}_i\fil{p}+\pd{r_j}\hat{v}_i\fil{\sigma'_{ij}}\\
&+\fil{\rho}(\lambda+\epsilon)-\hat{v}_i\pd{r_j}\hat{\tau}(v_i,v_j)\\
&+\fil{\rho}\hat{v}_i\hat{g}_i-\pd{r_j}\hat{\tau}(v_j,e_{\mathrm{int}})\,\,,
\end{split}\label{eq:filresetotsum}
\end{align}
where we introduced the total resolved
energy
$e_{\mathrm{res}}=\hat{e}_{\mathrm{int}}+\frac{1}{2}\hat{v_i}\hat{v_i}$,
being $\hat{e}_{\mathrm{int}}$ the filtered internal energy, and 
the generalized moments which are generically defined by
\begin{align}
\hat{\tau}(a,b)=&\ \fil{\rho a b} - \fil{\rho} \hat{a} \hat {b}\\
\begin{split}
\hat{\tau}(a,b,c)=&\ \fil{\rho a b c} - \fil{\rho} \hat{a} \hat{b} \hat{c} \\
&-\hat{a}\hat{\tau}(b,c)-\hat{b}\hat{\tau}(a,c)-\hat{c}\hat{\tau}(a,b)
\end{split}
\\
\hat{\tau}(a,b,c,d)=&\ \ldots
\end{align}
for Favre-filtered quantities $a$, $b$, $c$ etc. Germano interpreted the trace of 
$\hat{\tau}(v_i,v_j)\fil{\rho}$ 
as the squared velocity fluctuation, $q^{2}:=\hat{\tau}(v_i,v_i)/\fil{\rho}$.
The evolution of the corresponding turbulent energy, $e_{\mathrm{t}}=\frac{1}{2}q^{2}$,
is given by
\begin{align}
\pd{t}\fil{\rho}e_{\mathrm{t}}+\pd{r_j}\hat{v}_j
\fil{\rho}e_{\mathrm{t}}=\ \mathbb{D}+\Sigma+\Gamma-\fil{\rho}(\lambda+\epsilon)\,\,,
\label{eq:etsum}
\end{align}
where
\begin{align}
\mathbb{D} &= -\pd{r_j}\lrb{\frac{1}{2}\hat{\tau}(v_j,v_i,v_i) -\mu
+\kappa}\label{eq:def3}\\
\Sigma &= -\hat{\tau}_{ij} \partial v_i / \partial r_j \label{eq:def_prod}\\
\Gamma&=+\hat{\tau}(v_i,g_i)\\
\fil{\rho} \lambda &=\lrb{\fil{p}\pd{r_i}\hat{v}_i-\fil{p \pd{r_i}
v_i}}\label{eq:def4}\\
\fil{\rho} \epsilon
&=-\lrb{\fil{\sigma'_{ij}}\pd{r_j}\hat{v}_i-\fil{\sigma'_{ij} \pd{r_j} v_i}}\,\,,
\label{eq:def5}
\end{align}
and
\begin{align}
-\mu &=\fil{v_i p}-\hat{v}_i\fil{p}\\
-\kappa &=\fil{v_i \sigma'_{ij}}-\hat{v}_i\fil{\sigma'_{ij}}\,\,.
\end{align}
The explicit forms of the quantities $\mathbb{D},\lambda,\epsilon,\Gamma$ and
$\hat{\tau}(v_i,v_j)$ are unknown and have to be modeled in terms of closure
relations, i.~e., functions of the filtered flow quantities (or their derivatives) and the turbulent 
energy $e_{\mathrm{t}}$. The closures for all these terms represent the SGS model.

\subsection{Subgrid scale closures}
\label{closures}

In the following we consider a simplified set of equations to model
the influence of the turbulent small scale (SGS) motions on the
numerically resolved scales $\ell\ge\ell_\Delta$, neglecting the influence of the viscous
stress  
tensor $\fil{\sigma'_{ij}}$ (which is a very good approximation for high Reynolds numbers) and the 
turbulent transport of heat given by the divergence of
$\hat{\tau}(v_j,e_{\mathrm{int}})$ in equations~(\ref{eq:filmomsum})
and (\ref{eq:filresetotsum}). Moreover, gravitational effects 
on unresolved scales are neglected, i.~e., we set $\Gamma=0$ in equation~(\ref{eq:etsum}) for the
turbulent energy. For the terms~(\ref{eq:def3}), (\ref{eq:def_prod}),
and (\ref{eq:def5}), we adopt SGS closures that have been applied in
large eddy simulations of incompressible 
turbulence. The numerical study by \citet{Schmidt2006} demonstrated that these closures can
be carried over to transonic  turbulence, for which the unresolved
turbulent velocity fluctuations are small compared to the speed of
sound. Additionally, we utilize the pressure-dilatation model of
\citet{Sarkar1992} in order to account for moderate compressibility
effects. Since we concentrate on the 
dynamics of the gas  in the ICM, where the Mach numbers may locally approach unity, the
SGS closures outlined subsequently serve as a reasonable approximation. In supersonic
flow regions, on the other hand, the SGS model is deactivated in order to maintain
stability (see \S \ref{limits}).

The flux of kinetic energy from resolved scales toward subgrid scales, i.~e.,
the rate of turbulent energy production, is given by the contraction
of the turbulent stress tensor and the Jacobian of the resolved velocity field.
Since $\hat{\tau}_{kk}=\fil{\rho} q^2 $, we split the tensor in a symmetric trace-free part 
$\hat{\tau}^*_{ij}$ and a diagonal part:
\begin{align}
\hat{\tau}_{ij}=\hat{\tau}^*_{ij}+\frac{1}{3}\delta_{ij}\fil{\rho}q^2\,\,.
\end{align}

The model for $\hat{\tau}^*_{ij}$ is based on the turbulent viscosity
hypothesis \citep{Boussinesq1877}, which means
that $\hat{\tau}^*_{ij}$ is assumed to be of the same form as
the stress tensor $\sigma'_{ij}$ of a Newtonian fluid. Hence,
\begin{align}
\hat{\tau}^*_{ij}=- 2 \eta_{\mathrm{t}} S^*_{ij}
\end{align}
with a turbulent dynamic viscosity
$\eta_{\mathrm{t}}=\fil{\rho}\nu_{\mathrm{t}}=\fil{\rho}C_{\nu}l_{\Delta} q$ and
\begin{align}
S^*_{ij}=\frac{1}{2}\lra{\pd{r_j}\hat{v}_i+\pd{r_i}\hat{v}_j}
-\frac{1}{3}\delta_{ij}\pd{r_k}\hat{v}\,\,.
\end{align}
The turbulence production term is therefore modeled as
\begin{align}
\label{eq:turbvisc}
\hat{\tau}(v_i,v_j)=-2\fil{\rho}C_{\nu} l_{\Delta} q
S^*_{ij}+\frac{1}{3}\delta_{ij}\fil{\rho}q^2\,\,.
\end{align}
We set $C_{\nu}=0.05$ \citep{Sagaut2006}.

The SGS transport of turbulent energy (equation \ref{eq:def3}) is
modeled by a gradient-diffusion hypothesis, stating that the
non-linear term is proportional to the turbulent velocity $q^2$ 
gradient \citep[]{Sagaut2006}
\begin{align}
\label{eq:trans}
\mathbb{D}=\pd{r_i}C_{\mathbb{D}}\fil{\rho} l_{\Delta} q^2 \pd{r_i} q\,\,.
\end{align}
The diffusion coefficient has been calibrated to $C_{\mathbb{D}} \approx 0.4$ by
numerical experiments \citep{Schmidt2006}.

For sufficiently high Reynolds numbers, viscous energy dissipation (equation \ref{eq:def5}) becomes
entirely an SGS effect. The most simple
expression that can be built from the characteristic turbulent velocity and
length scale for dissipation is
\begin{align}
\epsilon=C_{\epsilon}\frac{q^3}{l_{\Delta}}\,\,. \label{eq:epsmodel}
\end{align}
For our simulations we  set $C_{\epsilon}=0.5$ \citep{Sagaut2006}.

The effect of unresolved pressure fluctuations in compressible turbulence is
described by the $\lambda$ term (equation \ref{eq:def4}). A simple
closure for subsonic turbulent flow is \citep{Dear73} 
\begin{align}
\lambda=C_{\lambda} q^2 \pd{r_i}\hat{v}_i\,\,,
\end{align}
where $C_{\lambda}=-\frac{1}{5}$ \citep{FurTab97}.
\citet{Sarkar1992} performed simulations of simple
compressible flows and investigated the influence of the mean Mach number of
the flow on the turbulent dissipation $\epsilon$ and the pressure dilatation
$\lambda$. Based on this analysis he suggested different models for these
terms, which we will describe in the following sections. These modifications
have been proven to yield good results for transonic turbulence \citep{Shyy1997}.

As a major effect of compressibility from direct numerical simulation,
\citet{Sarkar1992} identified that the growth rate of kinetic energy decreases
if the initial turbulent Mach number increases. This means that the
dissipation of kinetic energy (and, therefore, of the turbulent
energy) increases with the turbulent Mach number $M_{\mathrm{t}}=q/c_{\mathrm{s}}$,
where $c_{\mathrm{s}}$ is the speed of sound. 
\citet{Sarkar1992} suggested to account for this effect by using
\begin{align}
\label{sarkar1}
\epsilon=C_{\epsilon}\frac{q^3}{l_{\Delta}}(1+\alpha_1 M_{\mathrm{t}}^2)
\end{align}
with $\alpha_1=0.5$ as a model for the dissipation of turbulent energy.

Based on a decomposition of all variables of the equation for instantaneous
pressure
\begin{align}
\pdd{r_i} p = \pdd{t}\rho- 
\frac{\partial^2}{\partial r_i \partial r_j}(\rho v_i
v_j-\sigma'_{ij})\label{eq:instpres}
\end{align}
into a mean and a fluctuating part and comparisons with direct numerical 
simulations of simple compressible flows \citet{Sarkar1992} proposed a different
model for the pressure dilatation 
\begin{align}
\label{sarkar2}
\lambda=\alpha_2 M_{\mathrm{t}} \hat{\tau}^*_{ij} \ppd{r_j}{\hat{v}_i} 
- \alpha_3 M_{\mathrm{t}}^2 C_{\epsilon}\frac{q^3}{l_{\Delta}}
- 8 \alpha_4 M_{\mathrm{t}}^2 p_{\mathrm{t}} \ppd{r_k}{\hat{v}_k}
\end{align}
with $\alpha_2=0.15$,$\alpha_3=0.2$ obtained from a curve fit of the model with
direct numerical simulation (DNS). Unfortunately, \citet{Sarkar1992} does not specify a value for
$\alpha_4$, so there is some confusion in the literature about it. For example,
\citet{Shyy1997} set $\alpha_4 = 0$ and still found the Sarkar model in good
agreement with their DNS simulation. In this work, we adopt 
$\alpha_4 = \alpha_2^2/2$. With this choice, the effective production of
turbulent energy vanishes for a turbulent Mach number $M_{\mathrm{t}}=1/\alpha_2$.
In the following we will sometimes refer to the ``Sarkar SGS'' when
the equations (\ref{sarkar1}) and (\ref{sarkar2}) are used in the
model. 

\subsection{Filtered equations in comoving coordinates}

Simulations with a comoving cosmological background require a formulation of the filtered fluid 
dynamical equations in comoving coordinates. Applying the Germano decomposition
(\S \ref{decomp}) in a comoving coordinate system with spatially homogeneous scale
factor $a(t)$, we obtain\footnote{The comoving density $\tilde{\rho} = a^3 \rho$ and the 
comoving pressure $\tilde{p} = a^3 p$ are introduced to shorten the equations.}
\begin{align}
\pd{t}\fil{\tilde{\rho}} + \fa\pd{x_j}\hat{u}_j \fil{\tilde{\rho}} =&\ 0\,\,,
\label{eq:filcommass2}
\\
\begin{split}
\pd{t}\fil{\tilde{\rho}} \hat{u}_i 
+\fa\pd{x_j}\hat{u}_j \fil{\tilde{\rho}} \hat{u}_i =& 
-\fa\pd{x_i}\fil{\tilde{p}}+\fil{\tilde{\rho}}\hat{g}^*_i\\ 
&-\fa\pd{x_j}\hat{\tau}(u_i,u_j)- \fh \fil{\tilde{\rho}} \hat{u}_i\,\,,
\label{eq:filcommom2}
\end{split}
\\
\begin{split}
\pd{t}\fil{\tilde{\rho}}e_{\mathrm{res}}+\fa\pd{x_j}\hat{u}_j\fil{\tilde{\rho}}e_{\mathrm{res}}=&
-\fa\pd{x_i}\hat{u}_i\fil{\tilde{p}}
-\fa\fil{\tilde{\rho}}\hat{u}_i\hat{g}^*_i\\
&-\fh(\fil{\tilde{\rho}}e_{\mathrm{res}}
+\frac{1}{3}\fil{\tilde{\rho}}\hat{u}_i\hat{u}_i+\fil {\tilde{p}})\\
&+\fil{\tilde{\rho}}(\lambda+\epsilon)
-\fa\hat{u}_i\pd{x_j}\hat{\tau}(u_i,u_j)\label{eq:filcomresetot}
\end{split}
\\
\begin{split}
\pd{t}\fil{\tilde{\rho}}e_{\mathrm{t}}+\fa\pd{x_j}\hat{u}_j\fil{\tilde{\rho}}e_{\mathrm{t}}=&\
\mathbb{D} +\Gamma
-\fil{\tilde{\rho}}(\lambda+\epsilon)\\
&-\fa\hat{\tau}(u_j,u_i)\pd{x_j}\hat{u}_i
-2\fh\fil {\tilde{\rho}}e_{\mathrm{t}}\,\,.
\label{eq:filcomet}
\end{split}
\end{align}

With respect to the non-comoving equations listed in \S \ref{decomp}, the only
term to implement additionally is the last one on the right-hand side of equation 
(\ref{eq:filcomet}). Furthermore,
SGS closures have to be expressed in terms of the Jacobian of the velocity in comoving coordinates,
\begin{align}
J_{ij} = \pd{r_i}v_j&=\fa\pd{x_i}u_j + \fh\delta_{ij}.
\end{align}
In particular, the trace-free rate of strain tensor in comoving coordinates is given by
\begin{align}
S^*_{ij}=\frac{1}{2}\lra{J_{ij}+J_{ji}}-\frac{1}{3}\delta_{ij}J_{kk}\,\,.
\end{align}

\subsection{Limits of the SGS model}
\label{limits}
Numerical difficulties result from constraints on the validity of SGS closures.
In particular, the turbulent viscosity hypothesis expressed by equation (\ref{eq:turbvisc}) was devised to
account for the production of turbulence by shear in moderately compressible flow. This is typically
encountered in the
dense, central regions of galaxy clusters. However, the surrounding  low-density gas can be 
accelerated very quickly in the gravitational field of the cluster. Moreover, high velocity 
gradients are encountered in the vicinity of shocks which are produced by gas accretion onto 
filaments, sheets, and halos as well as by
the merging of substructures. In order to inhibit unphysical
production of turbulent energy by these mechanisms, which are not accommodated
in the present formulation of the SGS model, several numerical safeguarding mechanisms have been 
introduced. 

First of all, we implemented a simple shock detector which identifies strong negative divergence. 
A cell is marked if the velocity jump corresponding to the negative divergence
becomes greater than the sound speed across the cell width 
$l_{\Delta}$:
\begin{align}
\label{shock}
-\ppd{r_i}{v_i} > \frac{c_{\mathrm{s}}}{l_{\Delta}}\,\,.
\end{align}
In cells satisfying the above criterion, the source and transport terms of the 
SGS model (equations \ref{eq:turbvisc} and \ref{eq:trans}) are disabled. The turbulent energy is only
advected in these cells, and no coupling to the
the velocity and the resolved energy is applied. 

Besides the previous check, an additional constraint is imposed on the magnitude of the turbulent 
energy, via the turbulent Mach number. Basically,
the SGS model breaks down once $M_{\mathrm{t}}$ becomes large compared to unity, therefore a threshold for 
this quantity is set to $M_\mathrm{{t,max}}=\sqrt{2}$. This value for the maximal turbulent Mach number
is motivated by the theory of isothermal turbulence, where the effective gas
pressure can be expressed as
\begin{align}
p_{\mathrm{eff}}=\rho c_{\mathrm{s}}^2 + \frac{1}{3}\rho q^2 = \gamma_{\text{eff}} \rho c_{\mathrm{s}}^2\,\,,
\end{align}
and, consequently, $p_{\mathrm{eff}}$ is limited to the adiabatic value
$\frac{5}{3}\rho c_{\mathrm{s}}^2$. We verified that this threshold
does not harm our cluster simulations, because in the hotter gas
phases ($T > 10^5\ \mathrm{K}$) turbulence is largely subsonic, and
the threshold is rarely reached.   

A supplementary low temperature cutoff ensures that
the sound speed does not drop to excessively low values, which occur in cosmological simulations 
especially in the low-density voids. We set the lower limit of the temperature to 
$T_{\mathrm{min}}=\unit[10]{K}$. This threshold ensures numerical stability and does not affect 
the baryon physics appreciably, apart from possibly making the shocks on accreting structures weaker.

\subsection{Combining AMR and LES}
\label{amr+sgs}

In Large Eddy Simulations (LES), the filtered equations
(\ref{eq:filcommass2}-\ref{eq:filcomet}) are solved using an SGS model
as outlined in the previous Section.
However, the closure relations we use and, in fact, the very concept of
SGS turbulence energy only applies if the velocity fluctuations on subgrid scales are nearly
isotropic. This limits the LES methodology to flows where all
anisotropies stemming from large scale features, like boundary conditions or
external forces, can be resolved. In the {\sc fearless} method, the grid resolution $\ell_{\Delta}$
is locally adjusted by adaptive mesh refinement (AMR) in order to ensure that
the anisotropic, energy-containing scales are resolved everywhere. On the other hand,
it is assumed that turbulence is asymptotically isotropic on length scales
comparable to or less than $\ell_{\Delta}$. It is very difficult to justify the
latter assumption \emph{a priori}, because there are no refinement criteria
that would guarantee asymptotic isotropy on the smallest resolved length scales. By careful analysis
of simulation results, however, one can gain confidence that AMR resolves turbulent
regions appropriately.

As an infrastructure for the implementation of {\sc fearless}, we chose  
Enzo v.~1.0 \citep{oshea04}, an AMR, grid-based hybrid (hydrodynamics plus N-Body) code based on 
the PPM solver \citep{cw84} and especially designed for simulations of cosmological structure 
formation.
When a grid location is flagged for refinement in Enzo, a new finer grid is created, 
and the cell values on the finer grid are generated by interpolating them from the coarser grid 
using a conservative interpolation scheme. At each timestep of the coarse grid, the
values from the fine grid are averaged and the values computed on the 
coarse grid (in the region where fine and coarse grid overlap) are replaced. However,
this approach does not account for the inherent scale-dependence of the
turbulent energy. Assuming Kolmogorov scaling \citep{Kolmog41,Frisch1995}, the turbulent
energies at two different levels of refinement with cell size $l_{\Delta,1}$
and $l_{\Delta,2}$, respectively, are statistically related by
\begin{align}
\frac{e_{\mathrm{t},1}}{e_{\mathrm{t},2}}=\frac{q_1^2}{q_2^2}\sim
\lra{\frac{l_{\Delta,1}}{l_{\Delta,2}}}^{2/3}\,\,.
\label{eq:qscale}
\end{align}

Using this scaling relation, we implemented a simple algorithm
to adjust the turbulent energy budget when grids are refined or derefined.
The following procedure is used once a grid is refined:
\begin{enumerate}
\item Interpolate the values from the coarse to the fine grid using the standard
interpolation scheme from Enzo.
\item On the finer grid, correct the values of the
velocity components, $\hat{v}_i$, and the turbulent energy,
$e_{\mathrm{t}}=\frac{1}{2}q^2$, as follows
\begin{align}
\hat{v}'_i&=\hat{v}_i \sqrt{1+\frac{e_{\mathrm{t}}}{\hat{e}_{\mathrm{kin}}}\lra{1-r_{\Delta}^{-2/3}}},\\
e'_{\mathrm{t}}&=e_{\mathrm{t}} r_{\Delta}^{-2/3}\,\,,
\end{align}
where $\hat{e}_{\mathrm{kin}}$ is the resolved kinetic energy,
$r_{\Delta}$ is the refinement factor of the mesh, and the primed
quantities are the final values on the fine grid. The resolved 
energy is adjusted such that the sum of resolved energy and turbulent
energy remains conserved.
\end{enumerate}

Apart from adjusting the energy budget, the resolved flow should feature velocity
fluctuations on length scales smaller than the cutoff length of the parent grid if
a refined grid is generated. To address this problem we observe that the
smallest pre-existing eddies that are inherited from the parent grid 
will produce new eddies of smaller size within a turn-over time. Although
this implies a small delay because of the higher time resolution of the refined grid, 
the flow will rapidly adjust itself to the new grid resolution.

For grid derefinement we reverse this procedure:
\begin{enumerate}
\item Average the values from the fine grid and replace the corresponding values
on the coarse grid
\item In the regions of the coarse grid covered by finer grids,
correct the velocity components and the turbulent energy:
\begin{align}
e'_{\mathrm{t}}&=e_{\mathrm{t}}r_{\Delta}^{2/3},\\
\hat{v}'_i&=\hat{v}_i \sqrt{1-\frac{e_{\mathrm{t}}}{\hat{e}_{\mathrm{kin}}}\lra{r_{\Delta}^{2/3}-1}}\,\,.
\end{align}
Here, primed quantities denote the final values on the coarse grid. The resolved energy is
adjusted to maintain energy conservation and a positive kinetic energy.
\end{enumerate}

\section{Numerical tests}
\label{tests}

We applied two consistency tests of the SGS model in simulations of
forced isotropic turbulence in a periodic box. First, energy conservation was checked in
adiabatic turbulence simulations and, second, the scaling of the
turbulent energy over several levels of resolution was investigated
for isothermal turbulence. To simulate driven turbulent flow, a random forcing mechanism based
on the Ornstein-Uhlenbeck process was applied
\citep{Schmidt2004}. This  process generates a solenoidal (i.~e., divergence-free) stochastic force 
field which accelerates the fluid at large length scales $l\approx
l_{0}$, where $l_{0}$ is the size of the computational box. The
strength of the force 
field is characterized by a forcing Mach number $M_{\mathrm{f}}$. The Mach numbers of the flow
becomes comparable to $M_{\mathrm{f}}$ once the forcing has been applied over a
period of time that is defined by the integral time 
$t_{\mathrm{int}}=l_{0}/M_{\mathrm{f}}c_{\mathrm{s}}$.

\subsection{Energy conservation}
\label{numenergy}

For global energy conservation it turned out to be important
to compute the turbulent stress term in equation (\ref{eq:filmomsum})
indirectly as
\begin{align}
\hat{v}_i\pd{r_j}\hat{\tau}(v_i,v_j) = 
\pd{r_j}\hat{v}_i\hat{\tau}(v_i,v_j) - 
\hat{\tau}(v_j,v_i)\pd{r_j}\hat{v_i}\,\,,
\end{align}
The reason behind it is that only by using this rearrangement of the terms 
we can ensure that we do not introduce small numerical errors which would violate
the local sum 
\begin{align}
\pd{r_j}\hat{v}_i\hat{\tau}(v_i,v_j) =
\hat{v}_i\pd{r_j}\hat{\tau}(v_i,v_j)+\hat{\tau}(v_j,v_i)\pd{r_j}\hat{v_i}.
\end{align}
leading to a big error in global energy conservation.

As a testing case, we run a
LES of driven turbulence as outlined above on a static grid of $256^3$ grid points 
and periodic boundary conditions. The adiabatic index $\gamma=\frac{5}{3}$
and $M_{\mathrm{f}} = 0.68$. 

\begin{figure}[tp]
\centering
\includegraphics[width=0.9\linewidth]{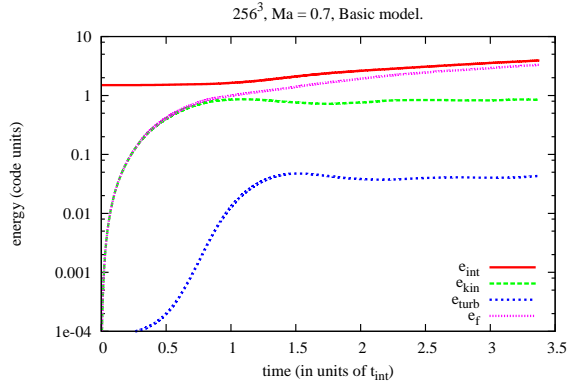}
\caption{Time evolution of mass-weighted energy averages in the driven
  turbulence simulation. The different energy components are indicated
  by colors.} 
\label{fig:sgs}
\end{figure}

Figure \ref{fig:sgs} shows the typical time development of
the mass-weighted mean energies in our simulation including the energy $e_{\mathrm{f}}$ injected
into the system by random forcing. It is evident from the curve of the turbulent
energy that after one integral time scale, our simulation reaches an equilibrium
between production and dissipation of turbulent energy.

\begin{figure}[tp]
\centering
\includegraphics[width=0.9\linewidth]{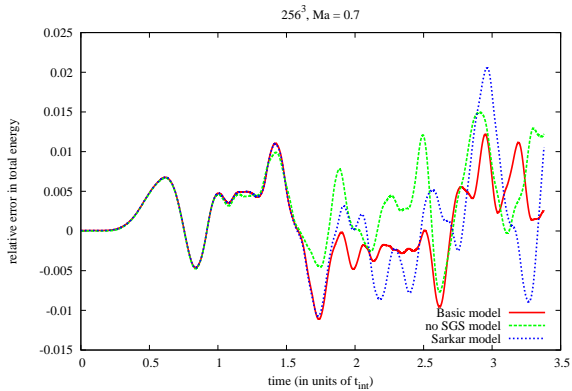}
\caption{Relative error of the total energy in the driven turbulence
  simulation. The different lines indicate simulations run with
  different versions of the SGS model, or without it, as shown in the
  legend.} 
\label{fig:relerror}
\end{figure}

In Fig.~\ref{fig:relerror}, we plot the time development of the relative
error $\frac{\Delta e(t)}{e(0)}$ of the mean total energy,
defined as the sum of the mass-weighted means of internal energy, kinetic energy,
turbulent energy minus the injected energy by the forcing,
\begin{align}
\hat{e}_{\mathrm{tot}}= \hat{e}_{\mathrm{int}}+\hat{e}_{\mathrm{kin}}+e_{\mathrm{t}}-\hat{e}_{\mathrm{f}}\,\,, 
\end{align}
where $\hat{e} = \frac{\fil{\rho e}}{\fil{\rho}}$. It demonstrates that with
our basic model, the relative error in energy is comparable to the error
without SGS model, and is around 1\%. The energy conservation of
the model using the Sarkar modifications is equally good. 

\begin{figure}[tp]
\centering
\includegraphics[width=0.9\linewidth]{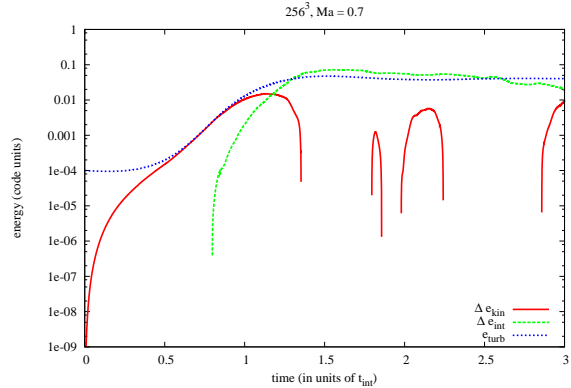}
\caption{Time evolution of the energy differences (red line: kinetic
  energy; green line: internal energy) between simulations with and
  without the SGS model (basic version), compared to the evolution of
  the turbulent energy (blue line).} 
\label{fig:diffenergy}
\end{figure}

It is also instructive to plot the difference between the energy contributions (internal and kinetic 
energy) in the 
simulations with and without the SGS model. These differences
are shown in Fig.~\ref{fig:diffenergy}. One can conclude from this figure
that, at the beginning of the turbulent driving, the turbulent energy produced
in our simulation with SGS model is found in the kinetic energy of the
simulation without SGS model. in contrast, from $t=1.2\ t_{\mathrm{int}}$ on, most of the turbulent energy
can be found in the internal energy of the simulation without SGS model.
Turbulent energy can therefore be interpreted as a kind of buffer which
prevents the kinetic energy in our simulation to be converted instantly into
thermal energy.  

\subsection{Scaling of turbulent energy}
\label{scalenergy}

\begin{figure}[tp]
\centering
\includegraphics[width=0.9\linewidth]{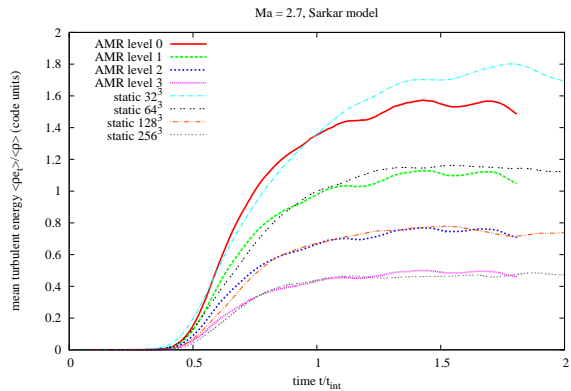}
\caption{Thick lines: mean mass-weighted turbulent energy for each level of the
 AMR simulation, using our procedure of transferring turbulent energy at grid
refinement/derefinement. Thin lines: the corresponding evolution of turbulent
energy of the static grid simulations. The colors indicate the AMR
level or the static grid resolution.} 
\label{fig:amrtrs2}
\end{figure}
\begin{figure}[tp]
\centering
\includegraphics[width=0.9\linewidth]{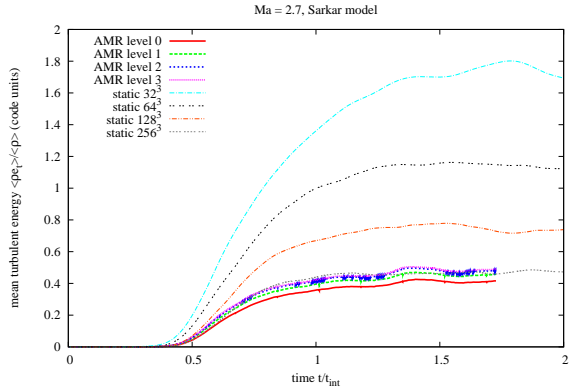}
\caption{Thick lines: mean mass-weighted turbulent energy for each level of the
 AMR simulation without transferring turbulent energy at grid 
refinement/derefinement. Thin lines: the corresponding evolution of turbulent
energy of the static grid simulations. The colors indicate the AMR level or the static grid resolution.}
\label{fig:amrtrs0}
\end{figure}

A necessary condition for the validity of the turbulent energy transfer algorithms
explained in \S \ref{amr+sgs} is that an AMR simulation should approximately
reproduce the results of static grid simulations corresponding to the different levels
of refinement. To test this, we compared
an AMR simulation of driven turbulence with a $32^3$ root grid resolution and three additional
levels (with a refinement factor of 2 between each level) to three static grid simulations with
resolutions of $32^3,64^3,128^3$ and $256^3$. In order to allow for the comparison
of averaged quantities, refinement of the entire domain was enforced at all levels
of the AMR simulation. In order to reach a statistically stationary root-mean-square
(rms) Mach number, we ran the simulations for nearly isothermal gas with $\gamma=1.01$
(for adiabatic turbulence, after an initial rise, the rms Mach number gradually decreases with time 
because of the dissipative heating of the gas).
We used a supersonic forcing Mach number, $M_{\mathrm{f}}=2.7$, to check whether
a consistent turbulent energy budget could be achieved for highly compressible
turbulence, albeit the scaling relations (\ref{eq:qscale}) of incompressible turbulence were 
utilized.

The results of this consistency check can be seen in Fig.~\ref{fig:amrtrs2}.
We observe that the time development of the mean turbulent energy is very similar on the
different levels of the AMR simulation compared to the static grid
simulations, except for some deviations at the root level. On the other hand, comparing these 
results to a simulation without correcting turbulent energy at grid
refinement/derefinement (Fig.~\ref{fig:amrtrs0}), it is evident that the scaling
of the turbulent energy in the latter case is inconsistent.

\section{Details of the cluster simulations}
\label{details}

\subsection{Simulation setup}
\label{setup}

We performed simulations of cluster formation with Enzo, following \citet{Iapichino2008}. 
We will compare two runs: one of them was done with the public version
of Enzo (without SGS model), and the other with {\sc fearless}
implemented, in its version including the Sarkar correction (equations
\ref{sarkar1} and \ref{sarkar2}).  
The simulations were done using a flat $\Lambda$CDM background cosmology with a dark energy density
$\Omega_{\Lambda}=0.7$, a total (including baryonic and dark matter) matter
density $\Omega_{m}=0.3$, a baryonic matter density $\Omega_{b}=0.04$,
the
Hubble parameter set to $\mathrm{h}=0.7$, the mass fluctuation amplitude $\sigma_8=0.9$,
and the scalar spectral index $n=1$. Both simulations were started with the same
initial conditions at redshift $z_{\mathrm{ini}}=60$, using the \citet{Eisenstein1999}
transfer function, and evolved to $z=0$. The simulations are adiabatic with a
heat capacity ratio $\gamma=5/3$ assuming a fully ionized gas with a mean
molecular weight $m_{\mu}= \unit[0.6]{u}$. Cooling physics, magnetic fields,
feedback, and transport processes are neglected. 

The simulation box has a comoving size of $\unit[128]{Mpc\ h^{-1}}$. It is
resolved with a root grid (level $l=0$) of $128^3$ cells and $128^3$ N-body
particles. A static child grid ($l=1$) is nested inside the root grid with a
size of $64\ \mathrm{Mpc\ h^{-1}}$, $128^3$ cells and $128^3$ N-body particles. The mass of each
particle in this grid is $\unit[9 \times 10^9]{M_{\odot}\ h^{-1}}$. Inside this grid,
in a volume of $\unit[38.4]{Mpc\ h^{-1}}$, adaptive grid refinement from level
$l=2$ to $l=7$ is enabled using an overdensity refinement criterion as
described in \citet{Iapichino2008} with an overdensity factor $f=4.0$. The refinement
factor between two levels was set to $r_{\Delta}=2$, allowing for an effective resolution
of $\unit[7.8]{kpc\ h^{-1}}$. 

The static and dynamically refined grids were nested around the place of
formation of a galaxy cluster, identified using the
HOP algorithm \citep{Eisenstein1998}. Since the realization of the
initial conditions was chosen identical to \citet{Iapichino2008}, this
study is based on the same cluster analyzed in that work. The cluster
has a virial mass of 
$M_{\mathrm{vir}}=\unit[5.95 \times 10^{14}]{M_{\odot}\ h^{-1}}$ and a virial radius of
$R_{\mathrm{vir}}=\unit[1.37]{Mpc\ h^{-1}}$.

\subsection{Local Kolmogorov scaling}
\label{local-scaling}

In static grid simulations one often chooses to use the grid resolution
$l_{\Delta}$ as characteristic length scale to compute a characteristic
velocity or eddy turnover time for this scale. However, in an AMR code it is
not trivial to compute the turbulent velocity $q_l$ associated with a characteristic
length scale $l = l_{\Delta}$, since $l_{\Delta}$ varies in time and space.
To circumvent this difficulty, we assume that below the grid
resolution turbulent velocity locally scales according to Kolmogorov
\begin{align}
q(l) \sim l^{1/3}\,\,.
\end{align}
We thereby assume that locally a Kolmogorov-like energy cascade sets in, at a length scale
given by the resolution of the grid at that position. This local hypothesis holds here only for 
the analysis of our simulations, and is similar to the assumption done in \S \ref{amr+sgs} for managing the grid refinement and derefinemnt. 

As a characteristic scale of our analysis,
we choose the length scale of our highest resolved regions, which is 
$l_{\mathrm{min}}=\unit[7.8]{kpc\ h^{-1}}$. The turbulent velocity in
the most finely
resolved regions can be computed directly from the values of the turbulent
energy $q(l) = \sqrt {2 e_{\mathrm{t}}}$ on the grid; the turbulent
velocity in less finely
refined regions is scaled down according to our local Kolmogorov hypothesis as
\begin{align}
q(l_{\mathrm{min}}) = q(l_{\Delta}) \lra{\frac{l_{\mathrm{min}}}{l_ {\Delta}}}^{2/3}
\label{eq:vturbscaled}.
\end{align}

\section{Results}
\label{results}

\subsection{Turbulent energy scaling in the cluster simulation}
\label{turbscaling}

\begin{figure}[tp]
\centering
\includegraphics[width=0.9\linewidth]{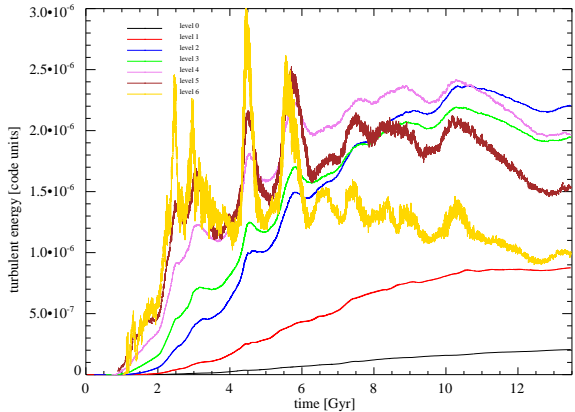}
\caption{Time evolution of the mean turbulent energy for each level of
refinement. This analysis has been
performed on a test run identical to that described in \S \ref{setup}
with only six AMR levels.} 
\label{fig:tuerescluster}
\end{figure}

In \S \ref{scalenergy}, we studied the temporal evolution of the turbulent energy
at different resolutions in a simulation of driven turbulence. In this
section, we repeat this analysis for our {\sc fearless} cluster
simulation. Figure \ref{fig:tuerescluster} shows the evolution
of the mass-weighted mean turbulent energy for every level of our AMR simulation. 
We see from the plot that the turbulent energy on the higher AMR levels $l$ (meaning at
smaller scales) is higher at early times ($2\ \mathrm{Gyr} < t < \unit[6]{Gyr}$). Later this picture
changes, but not completely. For example, the turbulent energy at $l = 4$ stays
above the turbulent energy at $l = 3$ throughout the
simulation. The magnitude of $e_{\mathrm{t}}$ along the AMR levels
suggests 
to locate the turbulence injection length scale between   
$125$ and $250\ \mathrm{kpc\ h^{-1}}$, corresponding to the effective
resolutions of levels 3 and 4. This is only a rough qualitative
estimate, but nevertheless in agreement with theoretical expectations
(e.g.~\citealt{ssh06}).   

Particularly noteworthy are the turbulent energy fluctuations on smaller
scales at the time $\unit[2]{Gyr}<t<\unit[6]{Gyr}$, corresponding to a redshift $z=3-1$. 
We can interpret these large
fluctuations as evidence for violent major mergers that happen at that
time, producing turbulent energy which is then dissipated into internal energy,
heating up the cluster gas. However, at $t > \unit[12]{Gyr}$ the
simulation seems to globally reach some kind of stable state,
comparable to what has been found in the driven turbulence
simulations.  

\subsection{Spatial distribution of turbulent energy}
\label{distribution}

\begin{figure*}
\centering
\includegraphics[width=0.9\linewidth,clip]{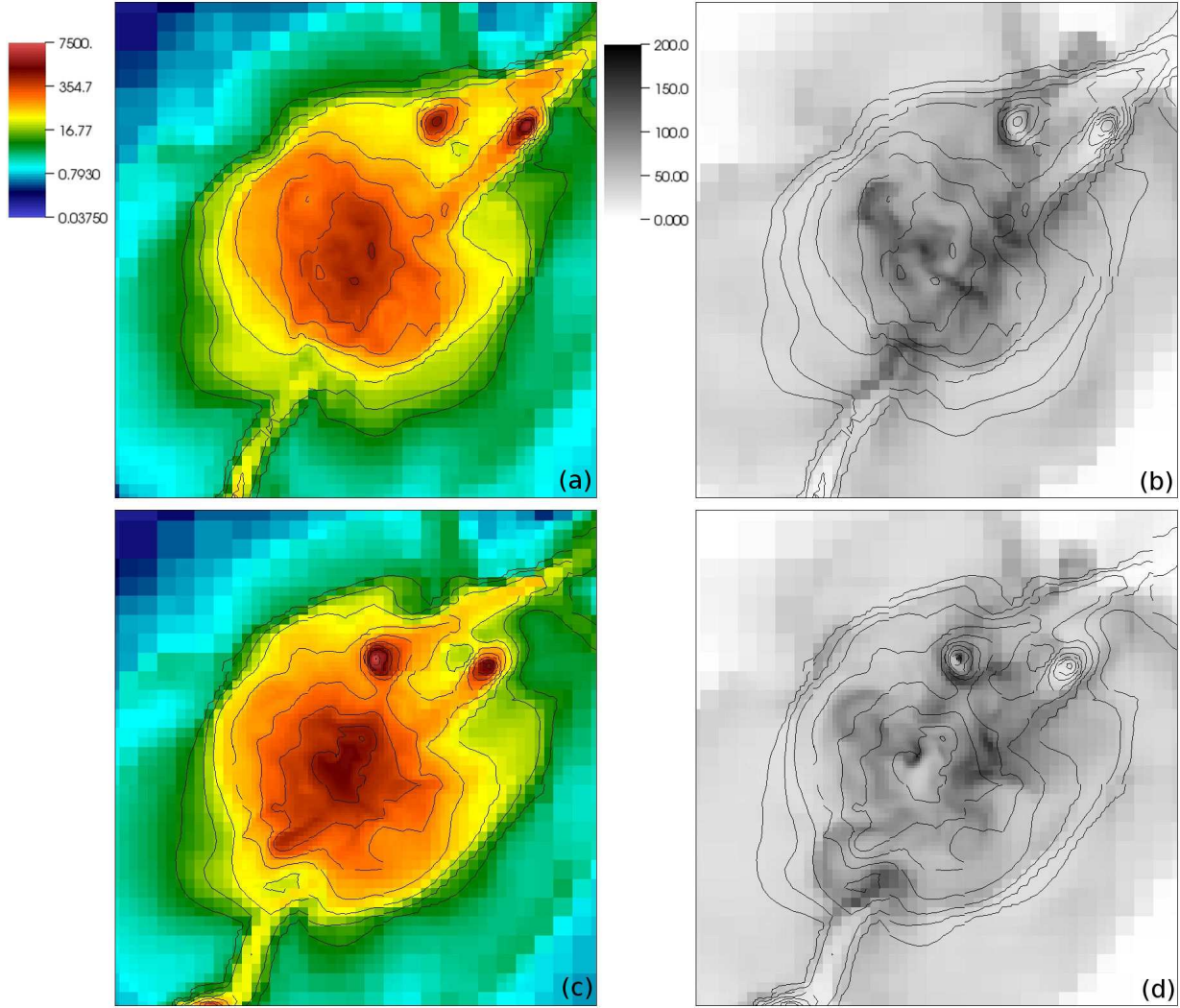}
\caption{Slices of baryon density (left-hand panels, $a$ and $c$) and
  turbulent velocity $q = \sqrt{2 e_{\mathrm{t}}}$ scaled to
$\unit[7.8] {kpc\ h^{-1}}$ (right-hand panels, $b$ and $d$) at
different redshifts $z$, for the {\sc fearless} run. The density is
logarithmically color coded as overdensity with respect to the average baryon density in the colorbar on the left of panel $a$, whereas $q$ is linearly coded
in $\mathrm{km\ s^{-1}}$, according to the colorbar on the left of panel $b$. The overlayed contours show density. 
The slices show a region of $\unit[6.4 \times 6.4] {Mpc\ h^{-1}}$
around the center of the main cluster followed in the
simulation. Panels $a$ and $b$ refer to $z = 0.15$, panels $c$ and $d$
to $z = 0.1$.} 
\label{slice-highz}
\end{figure*}

\begin{figure*}
\centering
\includegraphics[width=0.9\linewidth,clip]{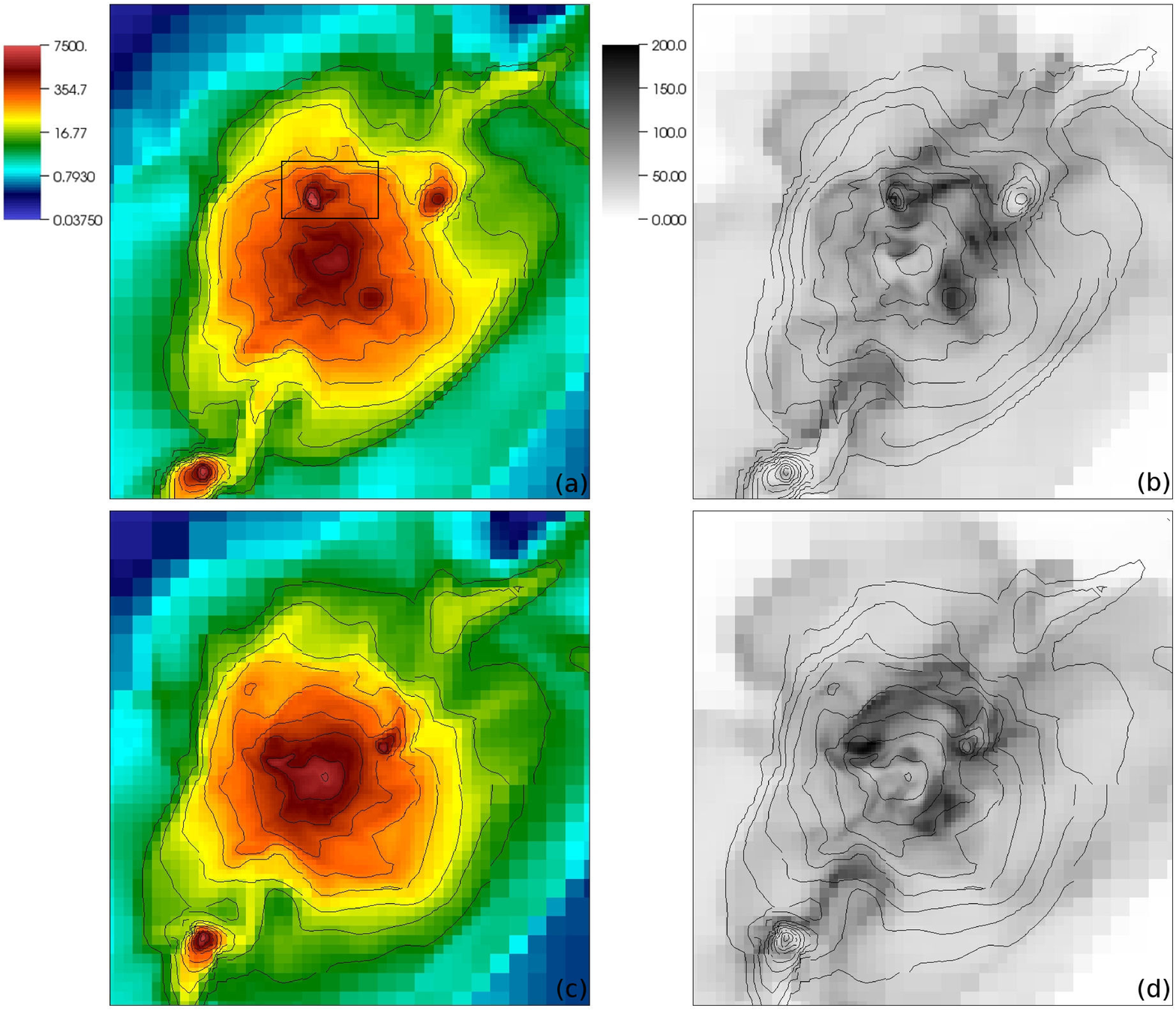}
\caption{Same as Fig.~\ref{slice-highz}, but in this case panels $a$
  and $b$ refer to $z = 0.05$, and panels $c$ and $d$ to $z = 0$. The black rectangle in panel $a$ denotes the projection on the slice of a small volume, including a subclump and its wake, analysed in Table \ref{subclump}.} 
\label{slice-lowz}
\end{figure*}

Before performing a quantitative analysis of the cluster properties in
the {\sc fearless} run, it is useful to visually inspect the
generation and the spatial distribution of the turbulent energy in the
ICM and around the cluster, in order to compare the simulations with
the theoretical expectations of cluster
mergers. Figs.~\ref{slice-highz} and \ref{slice-lowz} present a time
series of density and turbulent velocity slices, where several merger
events in the cluster outskirts can be identified. 

In the density slice at redshift $z=0.15$ (Fig.~\ref{slice-highz}$a$),
we can see a filament extending from the lower left to the upper right
corner of the figure; material is 
falling onto the cluster along this structure. On both sides, the
inflow of relatively cold gas from the filament onto the ICM produces
a moderate increase of turbulent energy (cf.~\citealt{nk03}). From the
upper right side there is not only a smooth inflow of matter, but two
small clumps are approaching the cluster. During the 
simulation these two clumps merge with the main cluster
(Figs.~\ref{slice-highz}$c$ to \ref{slice-lowz}$c$) and one of them
(on the left) is assimilated completely at 
redshift $z=0$. A substructure approaches the cluster from the lower
left corner along the filament (Fig.~\ref{slice-lowz}$a$) and another
one is visible only at $z = 0.05$ just to the right of the cluster
core, when it crosses the slicing plane.  

The merging process can be followed much more easily in terms of
production of turbulent energy, visualized by the turbulent velocity
$q = \sqrt{2 e_{\mathrm{t}}}$. 
In Fig.~\ref{slice-highz}$b$ at $z=0.15$, a marked peak of
turbulent energy in the center of our cluster, resulting from a former massive
merger, can be seen. The turbulent energy produced by this merger
declines (Figs.~\ref{slice-highz}$d$ to \ref{slice-lowz}$d$) and
at $z=0$ it is dissipated into internal energy nearly completely, as
confirmed by our further analysis in \S \ref{rad-profiles}.  

The two approaching clumps described above continue to drive turbulence in the cluster.
Thereby, the left clump can be identified in the turbulent velocity slice at
$z=0.15$ (Fig.~\ref{slice-highz}$b$) as a ring-like structure, showing that
turbulence is not produced at the center of the infalling clump but at the front (behind a 
bow shock) and in the wake of the infalling material. The right clump only
shows some turbulence production in its wake, which might be due to its
smaller size and smaller velocity. On their way towards the main cluster and through its ICM, both
clumps show a relevant production of turbulent energy
(Figs.~\ref{slice-highz}$d$ to \ref{slice-lowz}$b$). The considerable
amount of turbulent energy can 
even be identified after the two clumps have merged with the main cluster
(Fig.~\ref{slice-lowz}$d$) and the left one is not easily visible in
the corresponding density slice (Fig.~\ref{slice-lowz}$c$). From this
point of view, 
the distribution of turbulent energy traces the local merging history of a
galaxy cluster until it is dissipated into internal energy completely. 

The morphological evolution of the cluster gives a clear sense of the
markedly local behavior of the production and dissipation of
turbulence, which is confirmed to be an intermittent process in the
ICM.  

\subsection{Cluster energy budget}
\label{budget}

It is extremely difficult to apply an energy
analysis similar to that performed in \S
\ref{numenergy} to a galaxy cluster. Different than a periodic box, a galaxy cluster is an
open system, with a growth over time of negative gravitational
potential energy. Nevertheless, a comparison of the energy
contributions of the two simulations at $z = 0$ is useful to
understand the role of the SGS model. 

\begin{deluxetable}{cll}
\tablecaption{Energy contributions in a sphere of $r =
  R_{\mathrm{vir}}$, centered at the cluster core, at $z =
  0$. \label{cluster-values}} 
\startdata
\hline
\hline
\\
Quantity & Adiabatic run & {\sc fearless} run \\
\\
\hline
\\
$E_{\mathrm{tot}}\ [10^{63}\ \mathrm{erg}]$  & 2.6458  & 2.6426 (-0.1\%)   \\
$E_{\mathrm{int}}\ [10^{63}\ \mathrm{erg}]$  & 2.1982  & 2.2082 (+0.5\%)   \\
$E_{\mathrm{kin}}\ [10^{62}\ \mathrm{erg}]$  & 4.476   & 4.168  (-6.9\%)   \\
$E_{\mathrm{t}}\   [10^{61}\ \mathrm{erg}]$  & \nodata & 1.762             \\
\enddata
\tablecomments{The total energy $E_{\mathrm{tot}}$ is defined as the
  sum of $E_{\mathrm{int}}$, $E_{\mathrm{kin}}$ and, in the {\sc
    fearless} run, $E_{\mathrm{t}}$. The turbulent energy reported
  here is not scaled as described in \S \ref{turbscaling}.} 
\end{deluxetable}

The results are summarized in Table \ref{cluster-values}, where the
energies in a sphere centered at the cluster center and with $r =
R_{\mathrm{vir}}$ are reported. Different than elsewhere in this work,
in Table \ref{cluster-values} the energies are not specific,
i.e.~$E_{\mathrm{tot}} = \rho e_{\mathrm{tot}}$ etc. This choice
allows a better evaluation of the energy budget but it does not differ
appreciably from the analysis of specific energies, since the baryon
masses in the two runs agree within $1 \%$. 

The total energy remains basically unaltered in the two simulations,
whereas the most important change is the decrease of
$E_{\mathrm{kin}}$ in the {\sc fearless} run. The missing kinetic
energy is transferred mostly to $E_{\mathrm{t}}$, which acts
as a buffer between the resolved kinetic energy and
$E_{\mathrm{int}}$. The SGS model transfers energy from
$E_{\mathrm{t}}$ to $E_{\mathrm{int}}$ either adiabatically (via the
pressure dilatation term, equation \ref{eq:def4}) or irreversibly (via
the dissipative term, equation \ref{eq:def5}). Turbulent dissipation
is thus added to the dominant numerical dissipation, resulting in a
moderate increase of $E_{\mathrm{int}}$, though it is less relevant
than the variation of $E_{\mathrm{kin}}$.  

In both runs, the kinetic energy contribution in the cluster
$E_{\mathrm{kin}}$ is smaller than $E_{\mathrm{int}}$. The
mass-weighted average of the Mach number in the cluster is about 0.6,
in agreement with the known fact that the flow in the ICM is, on
average, mildly subsonic. In this regime it is not surprising that the
subgrid energy $E_{\mathrm{t}}$ is about two orders of magnitude
smaller than $E_{\mathrm{int}}$ on the global level. The energy
contribution from unresolved scales is globally negligible, though
locally turbulence can play a more significant role, as will be
discussed in \S \ref{rad-profiles}. 

Finally, we note that the ratio of the turbulent production term
$\Sigma$~(equation \ref{eq:def_prod}) to the turbulent dissipation
term $\epsilon$ (equation \ref{eq:def5}) in the cluster core is 
\begin{align}
\frac{\Sigma}{\epsilon} = 0.93
\end{align}
suggesting that the turbulent flow in the ICM is globally in a regime of near equilibrium of production and
dissipation of turbulent energy. 

\subsection{Radial profiles and local analyses}
\label{rad-profiles}

The results of \S \ref{budget}, referring to global features of the
galaxy cluster, will be complemented by a
local comparison in terms of radial profiles of selected physical
quantities and an analysis of the cluster core in this section.   

\begin{figure}
\plotone{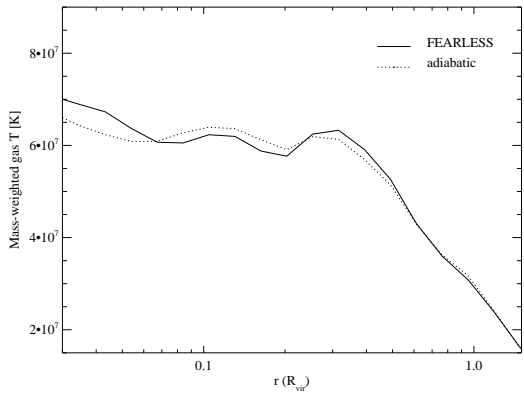}
\caption{Radial profiles of mass-weighted temperature at $z = 0$. The
  dotted line refers to the simulation without SGS model, whereas the
  solid line is for the {\sc fearless} simulation (in its version
  including the Sarkar corrections).} 
\label{fig:temp}
\end{figure}

\begin{figure}
\plotone{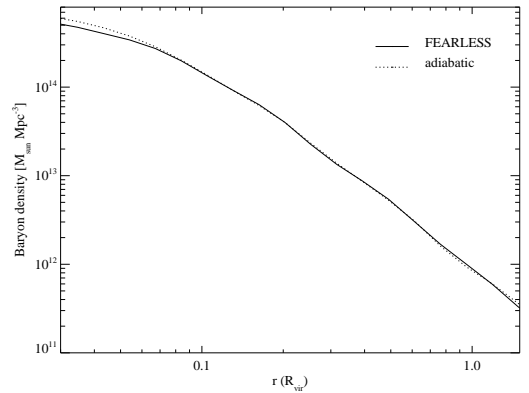}
\caption{Same as Fig.~\ref{fig:temp}, but showing the baryon density.}
\label{fig:dens}
\end{figure}

\begin{deluxetable}{cll}
\tablecaption{Mass-weighted averages in a sphere of $r = 0.07\
  R_{\mathrm{vir}}$, centered at the cluster core, at $z =
  0$. \label{core-values}} 
\startdata
\hline
\hline
\\
Quantity  & Adiabatic run & {\sc fearless} run \\
\\
\hline
\\
$\Sigma / \epsilon$                                           & \nodata & 0.59   \\
$e_{\mathrm{tot}}\ [10^{16}\ \mathrm{cm^2\ s^{-2}}]$          & 1.3781  & 1.4189 \\
$e_{\mathrm{int}}\ [10^{16}\ \mathrm{cm^2\ s^{-2}}]$          & 1.2529  & 1.3030 \\
$e_{\mathrm{kin}}\ [10^{15}\ \mathrm{cm^2\ s^{-2}}]$          & 1.252   & 1.138  \\
$e_{\mathrm{t}}\ [10^{13}\ \mathrm{cm^2\ s^{-2}}]$            & \nodata & 2.078  \\
$p_\mathrm{res} / (p_\mathrm{res} + p_\mathrm{therm})\ [\%]$  & 1.46    & 1.52   \\
$v_{\mathrm{rms}}\ [\mathrm{km\ s^{-1}}]$                     & 196     & 204    \\
\enddata
\end{deluxetable}

As a first interesting result of the comparison between radial
profiles in the {\sc fearless} simulation and in the standard
adiabatic run, one can notice (Fig.~\ref{fig:temp}) that the
temperature profile of the {\sc fearless} run deviates slightly from
that of the adiabatic run. This is especially apparent at the
center, where $T$ is larger in the cluster core for $r \lesssim 0.07\
R_{\mathrm{vir}}$, with respect to the standard run. Consequently
(Fig.~\ref{fig:dens}) the core in the {\sc fearless} run is less
dense, so that the ICM remains in hydrostatic equilibrium. The local
energy budget in the cluster core is therefore modified by the SGS
model. 

 In Table \ref{core-values}, we explore this feature in detail,
 reporting the mass-weighted averages of selected variables in a
 sphere within $0.07\ R_{\mathrm{vir}}$ from the cluster
 center. Because of the adjustment of the cluster hydrostatic
 equilibrium, the mass enclosed in this sphere is significantly
 different in the two runs (it decreases by $10\ \%$ in the {\sc
   fearless} run), thus it is more convenient to present specific
 energies in Table \ref{core-values}. 

First, the low value of the $\Sigma / \epsilon$ ratio indicates that,
at $z = 0$, in the cluster core region the dissipation of turbulence
is dominant with respect to its production. This confirms the
result of the morphological analysis in \S \ref{distribution}
that turbulence is not produced locally in the core by mergers at $z
< 0.15$, but that it decays in this region. The impact of
this turbulent dissipation on the local energy budget of the cluster
core can seen from the comparison of the energy contributions in Table
\ref{core-values}. Similar to the global analysis in Table
\ref{cluster-values}, there is a clear decrease of
$e_{\mathrm{kin}}$, transferred both to $e_{\mathrm{t}}$ and
$e_{\mathrm{int}}$. Both $e_{\mathrm{tot}}$ and $e_{\mathrm{int}}$ are
higher in the {\sc fearless} run, pointing to the existence of an
energy flux from the resolved scales to the thermal reservoir through
the turbulent buffer, leading to  the increase of the internal
energy. We interpret this additional energy contribution as caused by
the turbulent dissipation introduced by the SGS model. 

In the cluster core, the energy content at the subgrid scales is
marginal. Apparently the relative contribution to the total energy is
even smaller than in Table \ref{cluster-values}, but one should notice
that, for consistency, in that table both $E_{\mathrm{kin}}$ and
$E_{\mathrm{t}}$ are reported according to the original scale
separation introduced by the AMR resolution, and without rescaling
$E_{\mathrm{t}}$ as described in \S \ref{turbscaling}.  In the cluster
core the refinement level is maximum, therefore the unresolved part of
the turbulent cascade is relatively smaller than elsewhere, and so is
$e_{\mathrm{t}}$. In Table \ref{core-values} we use the scaled
definition of $e_{\mathrm{t}}$, but in the core it differs from the
unscaled one only marginally, because in this region the resolution is
$l_{\mathrm{min}}$ almost everywhere. 

\begin{figure}
\plotone{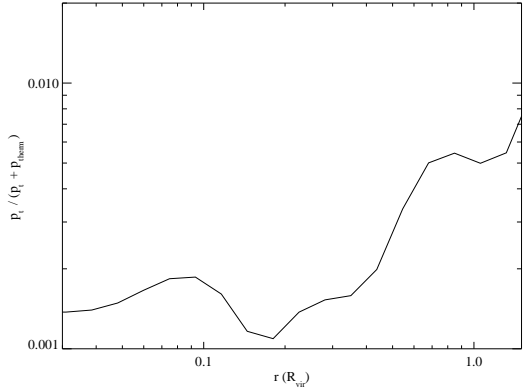}
\caption{Radial profile of the turbulent contribution to the pressure
  support $p_{\mathrm{t}}/(p_{\mathrm{t}}+p_{\mathrm{therm}})$, as
  defined in the text, for the {\sc fearless} simulation at $z = 0$.} 
\label{fig:pres}
\end{figure}

To further quantitatively appreciate the contribution of
$e_{\mathrm{t}}$ to the energy budget, Fig.~\ref{fig:pres} reports the
profile of the turbulent pressure support $p_\mathrm{t} /
(p_\mathrm{t} + p_\mathrm{therm})$ in the cluster, where  the
turbulent pressure is defined as $p_\mathrm{t} = 1/3\ \rho q^2$, and
$p_\mathrm{therm}$ is the usual thermodynamical pressure. This ratio
is also equal to the ratio of the corresponding energies
($e_\mathrm{t} / (e_\mathrm{t} + e_\mathrm{int})$). At the length
scale of the effective spatial resolution of the simulations
$l_{\mathrm{min}} = 7.8\ \mathrm{kpc\ h^{-1}}$, the contribution of
the turbulent pressure (or energy) is well below $1\%$, although it
increases at larger central distances. 

In analogy with the turbulent pressure, we define a ``resolved pressure''
\begin{align}
p_{\mathrm{res}}=\frac{1}{3} \rho v_{\mathrm{rms}}^2\,\,,
\end{align}
where the root-mean-square (hereafter rms) velocity \citep{Iapichino2008} is defined as
\begin{align}
v_{\mathrm{rms}} = \frac{\sum_i m_i \lra{v_i - \fil{v}}^2}{\sum_i m_i} \,\,.
\label{vrms}
\end{align}
Here, $\fil{v}$ is the mass-weighted average of the velocity in the
analysis volume. This quantity essentially probes the contribution of
turbulent motions at length scales of the order of $0.07\
R_{\mathrm{vir}} \sim 90\ \mathrm{kpc\ h^{-1}}$. As shown in Table
\ref{core-values}, the pressure contribution at these length scales is
at the percent level, and is slightly higher in the {\sc fearless}
simulation. Interestingly, the rms velocity in the {\sc fearless} run
is also somewhat larger than in the adiabatic case.   

\begin{figure}
\plotone{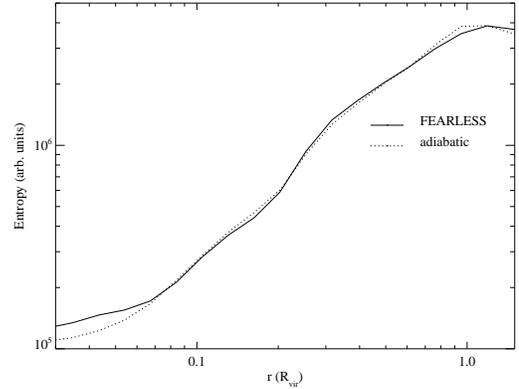}
\caption{Same as Fig.~\ref{fig:temp}, but showing the mass-weighted entropy (as defined in the text).}
\label{fig:entro}
\end{figure}

The changes in the temperature and density profiles are also reflected
on the entropy which is defined, as is customary in astrophysics, as 
\begin{align}
K=\frac{T}{\rho^{\gamma-1}}
\end{align}
with $\gamma = 5/3$. The entropy in the cluster core is higher in the
{\sc fearless} run as compared to the standard run
(Fig.~\ref{fig:entro}). This result is consistent both with the
locally increased dissipation of turbulent to internal energy provided
by the SGS model and with the higher degree of mixing induced in the
cluster core, shown by $v_{\mathrm{rms}}$ in Table \ref{core-values}. 

\begin{figure}
\plotone{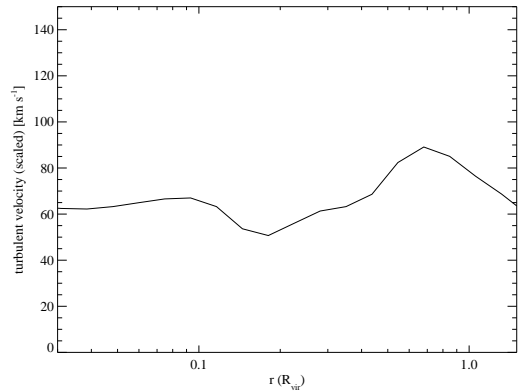}
\caption{Radial profile of the turbulent energy scaled at the length
  scale $l_{\mathrm{min}}$, as described in \S \ref{local-scaling},
  for the {\sc fearless} simulation at $z = 0$.} 
\label{fig:vturb}
\end{figure}

\begin{figure}
\plotone{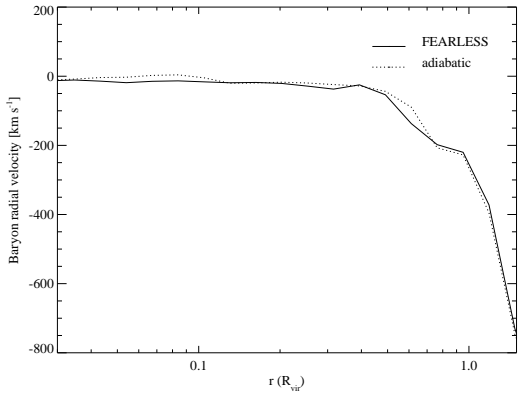}
\caption{Same as Fig.~\ref{fig:temp}, but showing the mass-weighted baryon radial velocity.}
\label{fig:vrad}
\end{figure}

The radial distribution of turbulent energy is displayed in
Fig.~\ref{fig:vturb} in terms of the turbulent velocity scaled to
$l_{\mathrm{min}}$. The turbulent velocity at this length scale is
below $100\ \mathrm{km\ s^{-1}}$. There is a pronounced peak at $r =
0.6\ R_{\mathrm{vir}}$ which is correlated with analogous trends in
the turbulent pressure (Fig.~\ref{fig:pres}) and in the radial
velocity profile (Fig.~\ref{fig:vrad}). This structure is clearly
linked with the most prominent merging clump shown in
Fig.~\ref{slice-lowz}$d$, ed analysed below in Table \ref{subclump}.  

There is an appealing similarity between the intervals of radii where
$q$ and the pressure ratio are larger ($r < 0.1\ R_{\mathrm{vir}}$
and $0.4\ R_{\mathrm{vir}} < r < 0.8\ R_{\mathrm{vir}}$), and the
corresponding intervals where the temperature and entropy of the 
{\sc fearless} run are slightly larger than those computed for the
adiabatic run. The opposite trend occurs in the interval in between,
where $v_{\mathrm{turb}}$ is comparatively smaller. The effects are
very small, but suggest that the SGS model plays the same
role in the ICM that was shown above for the cluster core, and in \S \ref{budget}
for the global quantities. In case of radial profiles, the spherical
averaging combined with the intermittent behavior of turbulence tends
to mask the turbulent effects. This can be better understood with a
comparison of the values of $q$ in Fig.~\ref{fig:vturb} and in the
right-hand panels of Figs.~\ref{slice-highz} and \ref{slice-lowz}: the
peak values in the slices are much larger than the spherical averages
in the profile. 

\begin{deluxetable}{cll}
\tablecaption{Mass-weighted averages in a volume of $[512 \times 768 \times 1280]\ \mathrm{kpc\ h^{-1}}$, containing a subclump and its wake, at $z = 0.05$. \label{subclump}} 
\startdata
\hline
\hline
\\
Quantity  & Adiabatic run & {\sc fearless} run \\
\\
\hline
\\
$\Sigma / \epsilon$                                           & \nodata & 1.13   \\
$e_{\mathrm{tot}}\ [10^{16}\ \mathrm{cm^2\ s^{-2}}]$          & 1.7447  & 1.5746 \\
$e_{\mathrm{int}}\ [10^{15}\ \mathrm{cm^2\ s^{-2}}]$          & 5.4281  & 5.9607 \\
$e_{\mathrm{kin}}\ [10^{16}\ \mathrm{cm^2\ s^{-2}}]$          & 1.2032  & 0.9786  \\
$e_{\mathrm{t}}\ [10^{14}\ \mathrm{cm^2\ s^{-2}}]$            & \nodata & 2.290  \\
$p_\mathrm{t} / (p_\mathrm{t} + p_\mathrm{therm})\ [\%]$      & \nodata & 3.70   \\
\enddata
\end{deluxetable}

The idea that locally the turbulence and its modeling can play a sizeable role is further corroborated by the data in Table \ref{subclump}, reporting the analysis at $z = 0.05$ of a small volume ($512 \times 768 \times 1280\ \mathrm{kpc\ h^{-1}}$) that contain one of the clumps presented in \S \ref{distribution} and its wake (cf.~Fig.~\ref{slice-lowz}$a$). The morphology of this accreting subcluster in the {\sc fearless} run is not substantially different from the the adiabatic one. The energy content, however, is rather different from that in the cluster core: in the region under consideration $e_{\mathrm{kin}}$ is dominant with respect to $e_{\mathrm{int}}$. The importance of the turbulence, injected by the hydrodynamical instabilities in the wake of the moving clump, is testified by the large ratio $\Sigma / \epsilon$ and by the turbulent pressure support, which is at the level of some percent, about one order of magnitude larger than the spherical averages in Fig.~\ref{fig:pres}. Despite of the slightly smaller average of $e_{\mathrm{tot}}$ in the {\sc fearless} run with respect to the adiabatic one, one can see an increase of $e_{\mathrm{int}}$, mostly at the expenses of $e_{\mathrm{kin}}$, resulting from the turbulent dissipation. The decrease of $e_{\mathrm{kin}}$ is rather large ($19\%$) but can be partially ascribed to the difficulty of comparing energy budgets in such open volumes. 

It is important to stress the deep difference between the turbulent
velocity profile in Fig.~\ref{fig:vturb} and the profile of
$v_{\mathrm{rms}}$, defined by equation (\ref{vrms}) (see also
\citealt{Norman1999a,Iapichino2008}). In the former case, the
mass-weighted average of a local quantity (i.e., defined in every cell)
is computed for each spherical shell, whereas in the latter
case $v_{\mathrm{rms}}$ is interpreted shell-wise as the standard
deviation with respect to the average $\fil{v}$. Clearly, the latter
definition does not retain any information related to a length scale,
and can be interpreted as turbulent velocity only in a loose
sense. From this point of view, the turbulent velocity provided by the
SGS model is a more powerful probe of the features of a turbulent
flow. On the other hand, spherically averaged velocity dispersions
(and the derived turbulent pressure) are meaningful in comparison with
observations, for example in the procedure for estimating the cluster
mass (cf.~\citealt{Rasia2006}). According to this different
definition, the spherically averaged turbulent pressure of the
simulated cluster (in a run similar to the adiabatic one presented
here) is reported in \citet{Iapichino2008}. It reaches values around
10\%, in agreement with the values found recently in simulations by
\citet{lkn09} for relaxed clusters. The turbulent pressure is somewhat
smaller in the {\sc fearless} run because of its slightly reduced
content in kinetic energy, but the difference is small, and is not
expected to significantly affect the estimates of the cluster mass. 

\section{Discussion and conclusions}
\label{conclusions}

Large Eddy Simulations (LES) are based on
the notion of 
filtering the fluid dynamic equations at a specific length scale, thus
performing a scale separation between the resolved and the unresolved
flow. The latter is treated by means of a subgrid scale model, which
in turn is coupled to the hydrodynamical equations governing the
former. In principle, a single scale separation is incompatible with
adaptive mesh refinement (AMR) codes, often used to study astrophysical phenomena. 

One of the aims of the present work was to address this numerical
problem by means of developing,
implementing, and applying a new numerical scheme that uses 
AMR and LES in combination, which we called {\sc fearless}. This novel tool is
suitable for modeling turbulent flows 
over a wide range of length scales, a key feature in the treatment of
many astrophysical flows including the intra-cluster medium. 

We showed that the idea of our approach to correct the
velocity and kinetic energy
at grid refinement/derefinement, according to local Kolmogorov scaling,
produces consistent results in simulations of driven turbulence. We demonstrated
that energy conservation and the scaling of turbulent energy in our adaptive
simulations is consistent with static grid simulations. 

To our knowledge, this work shows the first application of an SGS model
to simulations of the formation and evolution of a galaxy
cluster. The results give rise to several interesting implications
with regard to the physics of galaxy clusters and to the numerical methods
employed for their exploration in computational cosmology. 

The production of turbulence induced by minor mergers, analytically
studied by \citet{ssh06} and addressed by several numerical
investigations
\citep{hcf03,Takizawa2005,Takizawa2005b,afm07,dp08,ias08}, is
accurately tracked by the newly defined turbulent subgrid energy
(Figs.~\ref{slice-highz} and \ref{slice-lowz}), although the level of resolution of the idealised setups cannot be reached by cosmological simulations. The visualization and
subsequent analysis and postprocessing of turbulence and related
quantities is therefore easier and more consistent. Turbulence in the
ICM appears to be subsonic, in agreement with previous results. The
average ratio between the dissipation and the production term $\Sigma
/ \epsilon$ in the SGS model is close to unity, namely typical of a
system where the turbulence is roughly stationary. On the other hand,
this ratio is locally variable (cf.~Table \ref{core-values}) and,
together with the intermittent nature of turbulence in the ICM (\S
\ref{distribution}; see also \citealt{Iapichino2008}), delivers the
picture of a flow where turbulent motions are randomly initiated by
merger events and then gradually decay \citep{Frisch1995,ssh06}.  We
also notice that, in simulations of driven turbulence in a periodic
box (Fig.~\ref{fig:diffenergy}), the decrease of $e_{\mathrm{kin}}$
(noticed in our cluster simulation) is linked to an increase of
$e_{\mathrm{t}}$ only in the early driving phase, not in the later
equilibrium stage.  

The morphological evolution of the minor merger events and the
subsequent injection of turbulence in the ICM (Figs.~\ref{slice-highz}
and \ref{slice-lowz}) appear to be rather localized and intermittent,
confirming the feature of turbulent flows as being not very
volume-filling \citep{ssh06,Iapichino2008}. The dissipation of
turbulent to internal energy is thus modelled as a markedly local
process, consistent with the theoretical expectations.  

The effect of the SGS model on the cluster energy budget is well
exemplified by the comparison of our simulations at $z = 0$ (Table
\ref{cluster-values}). Although the value of $e_{\mathrm{t}}$ is small
compared to $e_{\mathrm{int}}$, this energy buffer is locally
effective in transferring the kinetic energy to the thermal
component. The dissipative effects are therefore more relevant in those
locations where $e_{\mathrm{t}}$ is relatively large, like the cluster
core (Fig.~\ref{fig:vturb}). In general, the main contribution of the
{\sc fearless} approach is to add a more physically motivated
contribution to the energy dissipation, which in Eulerian codes is
otherwise purely numerical. In {\sc fearless}, part of the energy flux
from resolved scales to the thermal reservoir is retained in the
buffer turbulent energy, $e_{\mathrm{t}}$, and is further dissipated
(turbulent dissipation) according to a local and temporal evolution
determined by the SGS model. 

Besides local effects, the importance of the SGS model for the overall
cluster structure appears small, because of the modest subgrid energy
contribution (Fig.~\ref{fig:pres}). One remark about
the simulated cluster is important at this point: as also verified in \citet{Iapichino2008}, this
structure is very relaxed (see also Fig.~\ref{fig:vrad}). Simulations
of more perturbed structures with recent or ongoing major mergers are
in preparation \citep{pim09}, because they will help to clarify the role of the
turbulent energy (and of its modelling) in the cluster energy budget
in cases where its magnitude is larger. From this viewpoint, the radial increase
of turbulent pressure support in the cluster outskirt
(Fig.~\ref{fig:pres}) is interesting for physical mechanisms (like the
acceleration of cosmic rays and magnetic field amplification) where
the knowledge of the turbulent state of the flow is needed.  
 
More turbulence in the cluster core is required, for example, to
reproduce the iron abundance profile in cool core clusters. 
Following \citet{dc05}, a turbulent diffusion coefficient can be
defined as $D_{\mathrm{turb}} \sim 0.1\ q\ l$, where $q$ is the
turbulent velocity at the length scale $l$. Using
$l = l_{\mathrm{min}}$ and $q \sim 60\ \mathrm{km\ s^{-1}}$, we find
$D_{\mathrm{turb}} \simeq 2\ \times 10^{28}\ \mathrm{cm^2\
  s^{-1}}$. We notice that this value is smaller than the estimates of
the effective diffusion coefficient in the cluster models of
\citet{Rebusco2005} and \citet{Rebusco2006}, which aim to reproduce
the turbulent diffusion of metals in the cores of selected
clusters. In particular, the cited models require much larger
turbulent velocities. In the framework of our cluster simulation,
these velocities could be injected into the ICM by a vigorous merger
event. Another possibility, explicitly suggested by the authors cited
above, is to invoke the action of an AGN outflow as an additional stirring agent in the cluster core.  

The enhanced temperature profile in the {\sc fearless} run is somehow
reminiscent of the theoretical predictions about the role of turbulent
heating in cluster cores \citep{dc05}. We notice an
apparent misunderstanding in the literature regarding this point. In
our model (and in the 
theory of turbulent flows in general), the dissipation of turbulent
energy does not act as an additional energy source but simply
releases the energy arising from the virialization process on a longer
timescale than the quick shock heating. Nevertheless, we showed that
turbulence, and the turbulent dissipation as well, can be rather
localized. Naively, one could think that an effective turbulent
heating in cool cores would require a peak of turbulent energy in the
cluster core, whose existence and magnitude should be justified
theoretically. Again, the stirring induced by AGN activity is an open
possibility which deserves further investigation. However, the model of
\citet{dc05} includes radiative cooling and thermal conduction, and a
detailed comparison is beyond the scope of the present work. We observe that additional physics which is here not addressed (thermal conduction, magnetic fields) could bring further interesting implications for the energy budget in the ICM and the turbulent mixing (\citealt{scq09}, and references therein).

Consequent to both the enhanced dissipation and fluid mixing is the
larger value of entropy in the {\sc fearless} cluster core. A
long-standing problem in cluster simulations is the shape of the entropy
profile, which smoothly decreases in the center in SPH
simulations whereas it flattens inside the core in runs with
grid-based codes \citep{fwb99}. This issue has been debated recently
in several works (among others, \citealt{Dolag2005}, \citealt{wvc08},
\citealt{koc09}), because it is controversial which difference of SPH
and mesh-based codes it results from. It has been claimed that the
source of discrepancy probably lies in the treatment of fluid mixing
\citep{mmb08}: the weaknesses of SPH in this regard are known, but the
ability of mesh codes to model the turbulent cascade on length
scales comparable with the grid resolution has not been addressed in a
satisfactory way. It is therefore unclear whether the flat core
entropy in grid codes correctly represents the physics of the ICM, or
perhaps numerical effects harm the robustness of this
feature. Recently, \citet{s09} pointed out that the core temperature
and entropy in grid-based codes are affected by a spurious increase,
caused by the N-body noise in the gravitational force field. 
In our opinion, the higher entropy core value in the {\sc fearless} run 
suggests that the typical flat entropy core is a hydrodynamical feature which requires a better understanding of the numerics in mesh codes, and is at least not primarily caused by N-body noise.

The SGS model applied in this work has to be considered as an
intermediate solution to address some basic questions related to
dynamics of the turbulent intra-cluster medium. A more elaborate model
that is able to handle the complexity of the flow (wide range of Mach
numbers and large density gradients as well as pronounced
inhomogeneities) in simulations of large scale structure evolution is
under development \citep{SchmFeder09}. This first application shows
the promising perspectives for the use of an SGS model in combination
with AMR and its potential impact on many branches of numerical
astrophysics. 

\acknowledgements
L.I.~acknowledges useful discussions with V.~An\-to\-nuc\-cio-Delogu,
M.~Bartelmann, S.~Borgani, G.~Murante, and M.~Viel.  
The simulations described in this work were performed using the Enzo
code, developed by the Laboratory for Computational Astrophysics at
the University of California in San Diego (http://lca.ucsd.edu). We
thank the Enzo development team, in particular D.~Collins, for
invaluable support in improving our modifications to the code. 
The numerical simulations were carried out on the SGI Altix 4700 \emph{HLRB2} of the Leibniz 
Computing Centre in Munich (Germany). 
The research of J.C.N. and L.I. was partly supported by the Alfried Krupp Prize for Young 
University Teachers of the Alfried Krupp von Bohlen und Halbach Foundation.

\bibliographystyle{apj}
\bibliography{Maier2009}

\end{document}